\documentclass[reprint,nofootinbib,amsmath,amssymb,fontenc aps]{revtex4-1}
\usepackage{graphicx,dcolumn,bm,hyperref,float}
\usepackage[mathlines]{lineno}

\newcommand{\beq}{\begin{equation}}
\newcommand{\eeq}{\end{equation}}
\newcommand{\bea}{\begin{eqnarray}}
\newcommand{\eea}{\end{eqnarray}}
\newcommand{\T}{\mathcal{T}}
\newcommand{\LGR}{\mathcal{L}_{\mbox{\tiny{GR}}}}
\newcommand{\Lkin}{\mathcal{L}_{\mbox{\tiny{kin}}}}
\newcommand{\Lint}{\mathcal{L}_{\mbox{\tiny{int}}}}
\newcommand{\Lex}{\mathcal{L}_{\mbox{\tiny{ex}}}}
\newcommand{\Lnonlocal}{\mathcal{L}_{\mbox{\tiny{non,local}}}}
\newcommand{\Lag}{\mathcal{L}}
\newcommand{\ks}{\tilde{\kappa}}
\newcommand{\rhor}{\rho_r}
\newcommand{\pri}{p_{r,i}}
\newcommand{\dpri}{\dot{p}_{r,i}}
\newcommand{\om}{\omega}
\newcommand{\M}{\mathcal{M}}
\newcommand{\f}{s_1}
\newcommand{\g}{s_2}
\newcommand{\mpl}{M_{\mbox{\tiny{Pl}}}}

\begin{document}
	
\title{Symmetries from Locality. II. Gravitation and Lorentz Boosts}
	
\author{Mark P.~Hertzberg$^1$}
\email{mark.hertzberg@tufts.edu}
\author{Jacob A.~Litterer$^1$}
\email{jacob.litterer@tufts.edu}
\author{McCullen Sandora$^2$}
\email{mccullen.sandora@gmail.com}
\affiliation{$^1$Institute of Cosmology, Department of Physics and Astronomy, Tufts University, Medford, MA 02155, USA
\looseness=-1}
\affiliation{$^2$Center for Particle Cosmology, Department of Physics \& Astronomy, 209 South 33rd Street, University of Pennsylvania, Philadelphia, PA 19104-6396, USA}

\date{\today}
	
\begin{abstract}
It is known that local, Lorentz invariant, unitary theories involving particles with spin 1 demand that the matter sector they couple to are organized by internal physical symmetries and the associated charge conservation, while spin 3/2 demands supersymmetry. However, the introduction of a spin 2 graviton does not obviously demand new symmetries of the matter sector (although it does demand a universal coupling). In this work we relax the assumption of Lorentz boost symmetry, while maintaining a basic notion of locality that there is no instantaneous signaling at a distance. This extends and complements our accompanying work in Part 1 on related issues for spin 1 particles in electromagnetism. In order to avoid potential problems with longitudinal modes of the graviton, we choose to project them out, leaving only two degrees of freedom. We study large classes of theories that a priori may violate Lorentz boost invariance. By requiring the tree-level exchange action be local, we find that consistency demands that the Lorentz boost symmetry must be satisfied by the graviton and the matter sector, and in turn we recover general relativity at this order of analysis.
\end{abstract}
	
\maketitle

\section{Introduction} 

The form of the basic interactions of nature are well known to be almost entirely dictated by the rules of quantum mechanics and special relativity, where the latter imposes (i) Lorentz symmetry and (ii) locality. In particular, there has been a large amount of work carried out over several decades demonstrating from various points of view the basic important conclusion: there is an essentially unique theory of a single type of massless spin 2 particles that is local, Lorentz invariant, and unitary with leading order interactions at large distances: it is general relativity (GR) \cite{Weinberg:1965rz,Deser:1969wk} (while multiple massless spin 2 with sub-leading interactions \cite{Wald:1986bj} can have problems with causality \cite{Hertzberg:2016djj,Hertzberg:2017abn}). Within the standard Lorentz invariant framework the only way to ``modify" gravity then is to introduce various types of new fields, especially new light scalar fields. However, the basic interactions of the graviton with itself and all other matter species is specified uniquely in terms of a universal coupling $G_N$ \cite{Weinberg:1964ew}, plus possible higher dimension operators that are unimportant at large distances (there can also be a cosmological constant, but that is not our focus here). 

While it is extremely powerful that Lorentz symmetry demands that the graviton must couple universally, one thing that we would like to note, however, is that at the classical level this does not place any additional restrictions on the matter sector. Let us elaborate on this as follows: Recall that when one includes a massless spin 1 particle into a theory and allows it to have leading order couplings to matter, then it must couple to an exactly conserved charge, associated with an internal ($U(1)$) symmetry. We emphasize that we are not referring to the (small) gauge symmetry, which is only a redundancy to remove the unphysical degrees in the field theory description of a massless spin 1 particle, but we are referring to the global sub-group of $U(1)$. This, by the Noether theorem, is associated with the conserved charge, and conversely, generates a symmetry that acts non-trivially on states. This actually imposes very significant constraints on the matter sector. For example, it forbids the matter sector coupled to it from being a single real scalar field; this cannot couple to a photon with leading order interactions. Similarly for multiple spin 1 we must impose some non-abelian symmetry on the matter sector and the spin 1 particles themselves. Again the consequences are significant; for instance it implies that a red quark must have a mass exactly equal to the mass of the blue quark; while this requirement would not be necessary in the absence of gluons. Furthermore, when massless spin 3/2 particles are introduced, they impose an even larger symmetry on the entire theory, namely that of supersymmetry. A summary of all this is provided in Table \ref{tableone}. 

\begin{table}[tb]
	\vskip.4cm
	\begin{center}
		\begin{tabular}{|c|c|c|c|}
			\hline 
			\bf{Particle} & \bf{Symmetry Demanded} \\
			\hline
			Spin 1 & Abelian ($U(1)$) \\
			\hline
			\,\,\,\,Multiple Spin 1\,\,\,\, & Non-abelian ($SU(N)$ etc) \\
			\hline
			Spin 3/2 & Supersymmetry \\
			\hline
			Spin 2 & \,\,\,\,? (This work: Lorentz boosts) \,\,\,\,\\
			\hline
		\end{tabular}
	\end{center}
\caption{{\em Assuming Lorentz boosts}: Different types of particles and the known corresponding physical symmetries that are needed in order for the matter sector to maintain Lorentz symmetry, locality, and unitarity. To be clear, we are referring to the physical {\em global} symmetries, which may be viewed as a special sub-group of the gauge group. We are not referring to the (small) gauge symmetries, which are just redundancies. In the last line, it is indicated that the spin 2 graviton is usually not  
known to enforce a new physical symmetry on the matter sector (it is sometimes said that it enforces general co-ordinate invariance, but this is another mere redundancy, and can be included simply by means of the Stueckelberg trick). \newline
{\em Without assuming Lorentz boosts}: Our accompanying work in Part 1 (Ref.~\cite{Accompanying}) showed that for a single spin 1 with 2 d.o.f.$~$we still need charge conservation (and hence the associated $U(1)$) in order to maintain locality. The primary goal of this work is to study the spin 2 case with 2 d.o.f.$~$and identify the required physical (global) symmetry from its consistency, which we explain are the Lorentz boosts themselves.}
\label{tableone}
\end{table}

On the other hand, if one considers some random matter sector, and then couples it to the graviton, one finds that (apart from possible gauge anomalies in chiral theories) there are {\em no additional constraints} imposed on the matter sector. For example, it can be trivially coupled to a real scalar or red and blue quarks with different masses, etc. (It does give rise to new BMS asymptotic symmetries \cite{BM,S1}, but this again does not restrict the matter sector.)
This may point to a missed understanding of the underlying reason for some symmetry in nature.
	
In this paper we would like to specify an underlying physical (global) symmetry that is in fact demanded of the graviton and matter sectors, which we usually just take for granted; this will be the Lorentz boost symmetry itself. This would be analogous to the following historical development: early work on quarks culminated in noting the need for the quarks to be organized by a  global $SU(3)$ symmetry in order to be compatible with some observations. However, the underlying origin of the $SU(3)$ symmetry remained obscure as it is easily deformable. Then after QCD's introduction of 8 massless interacting spin 1 particles, gluons, the color $SU(3)$ symmetry became demanded by consistency. 

Similarly, in this work, we suggest that while the Lorentz boost symmetry is easily deformable in the absence of the spin 2 graviton, it is demanded when the graviton is included. We emphasize that this is highly non-trivial and is highly non-standard. In contrast, it is sometimes said that GR is simply the theory that arises from ``gauging" the Poincar\'e symmetry to general co-ordinate invariance. But this is not meaningful, since general co-ordinate invariance can always be implemented trivially through the Stueckelberg trick. The key to gravitation is to actually {\em choose} to introduce the spin 2 graviton and then search for consistency with some over-arching principles. To make progress, we still need to invoke a very basic notion of locality, namely that there is no instantaneous action at a distance. 

We will start with GR and then introduce deformations of the Lorentz boost symmetry. We still maintain the idea of translation and rotation invariance in a preferred frame. The rotation invariance still allows us to organize particles by a notion of spin, and so we can again build theories of spin 2. This complements our accompanying work in Part 1 (Ref.~\cite{Accompanying}), in which we show how charge conservation is demanded in electromagnetism, merely from locality, although in that context Lorentz symmetry is easily deformable \cite{Colladay:1996iz,Colladay:1998fq,Coleman:1998ti}. This current work extends our earlier work in Ref.~\cite{Hertzberg:2017nzl}, where we also built the tree-level exchange action in a class of theories; but our present work will begin with a more general starting point and explore the possibilities more systematically. 
Other work on violating Lorentz invariance in the context of gravitation includes Refs.~\cite{AmelinoCamelia:2000mn,Magueijo:2001cr,Kostelecky:2003fs,Collins:2004bp,Mattingly:2005re,Horava:2009uw,Khoury:2013oqa,Collins:2006bw,Charmousis:2009tc,Papazoglou:2009fj,Griffin:2017wvh}.
	
Our paper is organized as follows:	
In Section \ref{GeneralG}, we recap GR and then formulate its Lorentz deforming generalizations.
In Sections \ref{grcase1}, \ref{grcase2}, \ref{grcase3}, and \ref{TT} we systematically study several classes of theories of spin 2 gravitons, and derive the consequences of locality on each of them.
In Section \ref{Discussion} we discuss and expand on our findings. 
Finally in the appendix we present supplementary material.

\section{General Relativity and its Generalization}\label{GeneralG}
	
Let us first discuss locality in the context of general relativity, before we discuss deformations of Lorentz boost symmetry. Starting with the full Einstein-Hilbert action (with vanishing cosmological constant), we can consider fluctuations of the spin 2 gravitational field $h_{\mu\nu}$ around a flat background (we use signature $+,-,-,-$ here) as follows
\beq
g_{\mu\nu} = \eta_{\mu\nu}+\kappa\,h_{\mu\nu}
\eeq
Then we obtain the quadratic Lagrangian density
\beq
\LGR =  \Lkin+\Lint
\eeq
where the graviton kinetic term is
\bea
\Lkin&=&\frac{1}{2} \left( \eta^{\alpha\beta} \partial_\alpha h^{\mu\nu} \partial_\beta h_{\mu\nu} - \eta^{\mu\nu} \partial_\mu h^{(4)} \partial_\nu h^{(4)} \right)\nonumber\\
&+&\partial_\mu h^{\mu\nu} \partial_\nu h^{(4)} - \partial_\mu h^{\mu\alpha} \partial_\nu h^\nu_\alpha 
\label{LGR}\eea
Here $h^{(4)}$ is the 4-dimensional trace $h^{(4)}=\eta^{\mu\nu} h_{\mu\nu}$. 
The leading order interaction is 
$\Lint =-{1\over2} \kappa\, h_{\mu\nu} T^{\mu\nu}$, 
where $T^{\mu\nu}$ is the energy-momentum tensor and the coupling $\kappa$ is related to Newton's gravitational constant $\kappa\equiv\sqrt{32\pi G_N}$. 

Ignoring the backreaction, $T^{\mu\nu}$ obeys the familiar equation of local energy and momentum conservation $\partial_\mu T^{\mu\nu}=0$. When broken up into components, this is
\bea
0&=&\partial_i T^{0i} + \dot T^{00}\label{ECON}\\
0&=&\partial_j T^{ij} +  \dot{T}^{0i}\label{PCON}
\eea

As is well known, general relativity avoids instantaneous action at a distance (as long as the source obeys the null energy condition $T_{\mu\nu}n^\mu n^\nu\geq0$). This can be seen clearly by operating in harmonic gauge $\partial_\mu h^\mu_\nu={1\over2}\partial_\nu h^{(4)}$. Then the equations of motion simplify to 
\beq
\square h_{\mu\nu}=-{\kappa\over2}\left(T_{\mu\nu}-{1\over2}T^{(4)} \eta_{\mu\nu}\right)
\eeq
(with $\square=\partial_t^2-\nabla^2$). 
As a basic test of locality we can compute the tree-level exchange action from picking up the particular solution $h_{\mu\nu}=-{\kappa\over2\square}\left(T_{\mu\nu}-{1\over2}T^{(4)} \eta_{\mu\nu}\right)$, which means we ignore external gravitons. 
The corresponding tree-level exchange action is half the interaction term $-{1\over4}\kappa\,h_{\mu\nu} T^{\mu\nu}$, giving the result
\bea
{8\Lex\over\kappa^2} &=& T_{\mu\nu}{T^{\mu\nu}\over\square}-{T^{(4)}\over2}{T^{(4)}\over\square}\\
&=&T_{ij} \frac{T_{ij}}{\square} - \frac{T}{2} \frac{T}{\square} + \frac{T_{00}}{2} \frac{T_{00}}{\square} - 2 T_{0i} \frac{T_{0i}}{\square} + T_{00} \frac{T}{\square} \,\,\,\,\,\,
\label{Lint_GR}\eea
where in the second line we have broken up $T^{\mu\nu}$ into components (with $T\equiv\delta_{ij}T^{ij}$). 
This result is clearly local, as it is given in terms of the inverse wave operator $\square$, the d'Alembertian, which has a retarded Green's function. In contrast, if we were to encounter inverse Laplacians, as we shall see later in the paper, this would imply instantaneous long ranged forces. This is a straightforward way to see that the most leading order processes in GR are local. When studying more general theories, examining the interaction Lagrangian this way provides a field theoretic procedure for diagnosing non-locality.
	
\subsection*{Generalization}

Our interest here is to not a priori assume boost invariance. However, we will still assume rotation invariance in a preferred frame. So it will be useful to decompose the field $h_{\mu\nu}$ into components that transform as a scalar, a vector, and a tensor under rotations, as follows
\beq
h_{00}\equiv \phi,\,\,\,\,\,\,\,\,h_{0i}=h_{i0}\equiv \psi_i,\,\,\,\,\,\,\,\, h_{ij}
\eeq
where $h_{ij}$ is associated with the polarizations of some spin 2 particle (graviton), and $\phi$ and $\psi_i$ are non-dynamical fields that will be useful tools to maintain locality (note that this $\phi$ is related to the Newtonian potential $\phi_N$ by $\phi=2\phi_N/\kappa$). We will use notation that the 3-trace is $h\equiv\delta^{ij} h_{ij}$. 

Now eq.~(\ref{LGR}) is comprised of dimension four terms in $\phi$, $\psi_i$, and $h_{ij}$, quadratic in fields and derivatives, with coefficients chosen to ensure Lorentz symmetry and propagate only 2 physical degrees of freedom. 
We will generalize the theory by inserting constant coefficients in front of every term that we denote $A,B,\ldots,L$, as follows
\bea
&\Lag&= -2A\,\dot{\psi}_i \partial_j h_{ij} + 2B\, \dot{h} \partial_i \psi_i  -C\, \partial_i \phi \partial_i h + D\, \partial_i \phi \partial_j h_{ij} \nonumber\\
&-& E\, \partial_i h \partial_j h_{ij} - F \left( \partial_i \psi_i \right)^2 + G\, \partial_j h_{ij} \partial_k h_{ik}  + H\,\partial_j \psi_i \partial_j \psi_i  \nonumber\\ &-& {I\over2} \dot{h}^2 + {J\over2} \partial_i h \partial_i h + {K\over2} \dot{h}_{ij} \dot{h}_{ij} - {L\over2} \partial_k h_{ij} \partial_k h_{ij}  + \Lint
\label{General}\eea
Note that the GR limit can be written, without loss of generality, as $A=B=\ldots=L=1$. There is one final quadratic spatial derivative term allowed by rotation invariance that could be included $\sim(\partial_i\phi)^2$ (without making $\phi$ or $\psi_i$ dynamical); this term is not part of GR and will not be the focus of most of the paper; however we will return to discuss it in Section \ref{TT}. Also, a mixed temporal-spatial derivative term $\sim\dot\phi\,\partial_i\psi_i$ could be added, but can be readily shown to lead to non-locality, and so it will be ignored here. While quadratic temporal derivative terms $\sim(\dot\phi)^2,\,\dot\phi\,\dot{h},\,(\dot\psi_i)^2$ lead to additional degrees of freedom (which is not our focus; see next subsection) and so will be set to zero. 

One could also include Lorentz violating mass terms into this action (see \cite{Dubovsky:2004sg}), although current data suggests the graviton is massless \cite{Tanabashi:2018oca}. Furthermore, as we will mention in the next subsection, by cutting down to 2 degrees of freedom, mass terms generally lead to non-locality. In the first upcoming theory in Section \ref{grcase1}, we will go through the details of mass terms in a subsection to in fact show that this is the case. However, for simplicity, we will not go through the full details in later sections. Finally, single derivative terms can be eliminated by using the mass terms and field re-definitions, generating dimension 5 operators that we ignore.

In this more general case we write the interaction using the notation
\beq
\Lint =-{1\over2} \kappa \, h_{\mu\nu} \T^{\mu\nu}
\eeq
where $\T^{\mu\nu}$ does not \emph{a priori} have anything to do with the conserved energy-momentum tensor $T^{\mu\nu}$. We can decompose it into its scalar, vector, and tensor pieces, which we denote as
\beq
\T^{00}\equiv \rho,  ~~~~ \T^{0i} = \T^{i0} \equiv p_i, ~~~~ \T^{ij} \equiv \tau_{ij}
\label{Tcomponents}\eeq
and we emphasize that \emph{a priori} they need not be related to energy density, momentum density, pressure or stress (however, those connections will emerge later in the paper). Note that more coefficients inserted in the interaction terms could simply be absorbed into the sources by a redefinition without loss of generality. In terms of parameters, we have included a total of 12 new parameters ($A,B,\ldots,L$), in addition to the coupling strength of gravity $\kappa$. However, we have yet to canonically normalize our fields. We have the freedom to re-scale our fields $\phi$, $\psi_i$ and $h_{ij}$, to eliminate 3 parameters, as well as to re-scale our units to set the graviton speed $L$ to unity. So in fact we have at most $12-4=8$ physical parameters that characterize non-Lorentz invariant deformations of general relativity. 

There is one further consideration: in this framework the trace of $\tau_{ij}$ is another scalar, which exhibits some partial degeneracy with the scalar source $\rho$. For the gravitational field, there is a similar degeneracy in the meaning of the scalars $\phi$ and the trace of $h_{ij}$, i.e., $h$. We can use this to eliminate one more parameter in the action (\ref{General}). In particular, consider the following pair of transformations of the sources and fields
\begin{eqnarray}
&&\tau_{ij}\to\tau_{ij}+\left(E-J\over 2C-D\right)\delta_{ij}\,\rho\label{tauTF}\\
&&\phi\to\phi-\left(E-J\over 2C-D\right)h\label{phiTF}
\end{eqnarray}
This leaves $h_{\mu\nu}\mathcal{T}^{\mu\nu}$ unchanged. Furthermore, the structural form of the starting action eq.~(\ref{General}) is unchanged, except, one has now mapped the coefficient of the $\partial_ih\partial_j h_{ij}$ term ($E\to\bar{E}$) and the coefficient of the $\partial_ih\partial_ih$ term ($J\to\bar{J}$) to be equal to one another $\bar{E}=\bar{J}=(2CE-DJ)/(2C-D)$, which removes another parameter. This, along with the ability to rescale the 3 kinds of fields and to set units for length vs time, means the remaining number of parameters is $12-4-1=7$.

Varying the Lagrangian in eq.~(\ref{General}) gives the equations of motion 
\bea
\ks \, \rho &=& -C \nabla^2 h + D \partial_i \partial_j h_{ij} \label{equations of motion1} \\
\ks \, p_i &=&  B \partial_i \dot{h} + H \nabla^2 \psi_i -F \partial_i \partial_j \psi_j- A \partial_j \dot{h}_{ij} \label{equations of motion2} \\
\ks \, \tau_{ij} &=&  2B \delta_{ij} \partial_k \dot{\psi}_k - I \delta_{ij} \ddot{h} + K \ddot{h}_{ij} -A \partial_{(i} \dot{\psi}_{j)}   \nonumber\\
&+& D \partial_i \partial_j \phi - E \partial_i \partial_j h - E \delta_{ij} \partial_k \partial_l h_{kl} - C \delta_{ij} \nabla^2 \phi  \nonumber\\
&+& G \partial_k \partial_{(i}h_{j)k}+J \delta_{ij} \nabla^2 h - L \nabla^2 h_{ij}.
\label{equations of motion}
\eea
where $\ks\equiv-\kappa/2$.  

In general we do not need to necessarily have conservation of $\T^{\mu\nu}$ in this framework, as we do in GR, given in eqs.~(\ref{ECON},\ref{PCON}). We parameterize the breakdown of local source conservation by functions $\sigma$ and $w_i$ as follows
\bea
\sigma &\equiv& \partial_i p_i + \dot{\rhor} \\
w_i &\equiv& \partial_j \tau_{ij} +  \dpri 
\eea
where $\rhor\equiv(A/D)\,\rho$ and $\pri\equiv(A/H)\,p_i$ are conveniently re-scaled densities. When assuming Lorentz symmetry, as is the underlying symmetry of general relativity, we know $\T^{\mu\nu}\to T^{\mu\nu}$, whose conservation $\partial_\mu T^{\mu\nu}=0$ implies $\sigma=w_i=0$. But in general these may be non-zero in our much larger class of theories that arise from not assuming Lorentz boost symmetry. We can now use the equations of motion to determine these possible violations of source conservation as follows
\bea
&\ks& \sigma =  \left( B - \frac{A C}{D} \right) \nabla^2 \dot{h}  + (H-F) \nabla^2 \partial_i \psi_i\label{sigmaEq}\\
&\ks& w_i = \left( 2B - A - \frac{AF}{H} \right) \partial_i \partial_j \dot{\psi}_j + \left( \frac{AB}{H} - I \right) \partial_i \ddot{h} \nonumber\\
&+& \left( J - E \right) \partial_i \nabla^2 h +  \left( G - L \right) \nabla^2 \partial_j h_{ij} + \left(D - C \right) \partial_i \nabla^2 \phi\nonumber\\
&+& \left( K - \frac{A^2}{H} \right) \partial_j \ddot{h}_{ij} + \left( G - E \right) \partial_i \partial_j \partial_k h_{jk}\,\,\label{wEq}
\eea
Note that in the GR limit all parameters can be set to $A=B=\ldots=L=1$ and so every term on the right hand side of this pair of equations vanishes, ensuring conservation of sources.

\subsection*{Degrees of Freedom}
	
The Lagrangian density in eq.~(\ref{LGR}) describes a graviton with two degrees of freedom (helicities), which is ensured by the presence of the familiar gauge redundancy: $h_{\mu\nu}\to h_{\mu\nu}+\partial_{(\mu}\alpha_{\nu)}$ and the presence of constraints. This is ordinarily understood as needed to describe the massless spin 2 representation of the Lorentz group. In this work we will deform away from Lorentz symmetry and then the number of degrees of freedom is less clear. However, we take the following guide: In the Lorentz invariant case, the only way to have more than 2 degrees of freedom is to make the graviton massive, which then introduces a total of five degrees of freedom. However, the most longitudinal modes appear to exhibit  strong coupling problems at short distances. In the ghost free versions of massive gravity \cite{deRham:2010kj}, this occurs at the rather low scale of $\Lambda\sim(m^2\mpl)^{1/3}$, where $m$ is the graviton mass and $\mpl=1/\sqrt{G_N}$ is the Planck mass \cite{ArkaniHamed:2002sp}. For example, if the graviton mass $m\sim H_0$, where $H_0\sim 10^{-33}$\,eV is today's Hubble parameter, then this strong coupling scale is the rather low $\Lambda\sim 10^{-13}$\,eV. It is generally believed that this requires a UV completion at this rather low scale in terms of the massless theory anyhow, returning then to just 2 fundamental degrees of freedom in the graviton sector. Also, it has been argued by some these longitudinal modes cause serious problems for the consistency of even the low energy effective theory; that there are potential problems from acausality \cite{Adams:2006sv,Deser:2012qx,Deser:2014hga}, although this is an ongoing discussion (for a review, see Ref.~\cite{deRham:2014zqa}).

In our case of interest, we are deforming away from the Lorentz symmetry. In this case it is less clear that the above problems of strong coupling and/or acausality of the longitudinal modes persist. However, we can focus our attention in this work on small deformations of the Lorentz symmetry, as we did in our accompanying work \cite{Accompanying}. In this case the potential problems with more degrees of freedom are still a concern, so the most direct approach to avoid such problems is to project out those additional degrees of freedom, and still only build theories of the 2 helicities of the graviton. Once this projection is made, it is relatively straightforward to show that action at a distance would occur if the graviton's mass were non-zero, as we will show in a subsection of Section \ref{grcase1} (similar to what we showed explicitly in Ref.~\cite{Accompanying} in the analogous electromagnetic case). So for the most part in this work, we shall set the mass to zero. We note that recent measurements of gravitational waves by LIGO \cite{Abbott:2016blz} are  consistent with zero mass for the graviton as well as just 2 propagating modes. While this is not a proof that such additional modes don't exist, it does provide further motivation for our setup.

The generalized Lagrangian eq.~(\ref{General}) breaks Lorentz boost invariance, and now appears to contain 12 new free parameters (though some could be eliminated by rescaling the fields), and without imposing the gauge redundancy of GR, carries additional degrees of freedom. Since we wish to study theories with only two degrees of freedom, we must put some constraints on the fields or parameters to cut down to the desired two propagating modes. There appear to be few ways to accomplish this: (A) Set $\partial_i \psi_i = 0 = \partial_j h_{ij}$, for which the equations of motion imply the sources are conserved (or at least a shifted version is); (B) Fix parameters such that $\partial_i \psi_i$ and  $\partial_j h_{ij}$ are directly determined by the equations of motion; (C) Set $\partial_i \psi_i = \mathcal{F}_0$, $\partial_j h_{ij} = \mathcal{F}_i$, where $\mathcal{F}_\mu=\mathcal{F}_\mu[\Psi_m]$ are some non-dynamical functions of the matter fields $\Psi_m$ and constructed to be consistent with the equations of motion; (D) include an additional term $\sim(\partial_i\phi)^2$ in the action and demand the field is transverse-traceless $\partial_i h_{ij}=h=0$. In each case, we shall build the most general theory compatible with these starting points and then demand the tree-level exchange action be local.

Since we are not pre-supposing Lorentz symmetry, we don't need to invoke gauge redundancy of GR in order to describe physics in a manifestly local way. So any restriction on the fields is not simply a ``gauge choice,'' but a choice of theory; choices (A), (B), (C), and (D) are different theories of spin 2 with two degrees of freedom (although (A), (B), (C) will be connected to each other once we restrict their parameters). We neither assume anything about what physical quantities $\T^{\mu\nu}$ represents, only than it should be a non-trivial source and not overly constrained -- it is simply the symmetric source which couples to  $h_{\mu\nu}$. Only \textit{a posteriori} will we be able to identify the form that $\T^{\mu\nu}$ must take.

\section{Theory A: Transverse Constraint}\label{grcase1}	
	
We wish to apply essentially the same procedure as we did with spin 1 \cite{Accompanying}. We want to require the theory of eq.~(\ref{General}) be local and see whether we recover an interaction like eq.~(\ref{Lint_GR}). In this section, we do this by cutting down to two degrees of freedom by requiring the divergences of the fields vanish. 
	
In this section we directly remove any longitudinal components of the gravitational fields. It is the analogue of the Coulomb constraint in electromagnetism ($\partial_i A_i=0$)
\beq
\partial_i \psi_i = 0,\,\,\,\,\,\,\,\,\,\,\partial_j h_{ij} = 0  \label{gauge}
\eeq
Since we are not working in GR, we do not need to carry around any gauge redundancy to make our assumed space-time symmetries manifest (as we only wish to make rotation invariance manifest, while gauge redundancy is useful to make Lorentz symmetry manifest). So this constitutes a choice of theory. In fact it may be implemented by using a pair of Lagrange multipliers in the action. In this theory, the equations of motion become
\bea
\ks \, \rho &=& -C \nabla^2 h  \label{rhoA}\\
\ks \, p_i &=& B \partial_i \dot{h} + H \nabla^2 \psi_i  \\
\ks \, \tau_{ij} &=&  - I \delta_{ij} \ddot{h} + K \ddot{h}_{ij} -A \partial_{(i} \dot{\psi}_{j)} - C \delta_{ij} \nabla^2 \phi \nonumber\label{pA}\\
&+& D \partial_i \partial_j \phi - E \partial_i \partial_j h  + J \delta_{ij} \nabla^2 h - L \nabla^2 h_{ij}\,\,\,\,\label{tauA}
\eea
Note the parameters $F$ and $G$ no longer appear; the choice eq.~(\ref{gauge}) removes these terms from the (classical) theory. Hence we are now down to $12-4-1-2=5$ physical parameters relevant to our analysis.
	
The usual statement of conservation of sources is $\partial_\mu T^{\mu\nu}=0$. 
Using the above equations of motion in this theory, we can write the zeroth component as
\beq
\partial_i p_i + \dot{\rho}_a = 0 \label{cons0}
\eeq
which is of the usual form ($\rho_a\equiv(B/C)\,\rho$ is a simple re-scaling of $\rho$, which is just a matter of convention). Checking the other three components, these do not immediately form a canonical conservation equation like eq.~(\ref{cons0}):
\bea
\partial_j \tau_{ij} + \dpri &=& \left( I - \frac{AB}{H} \right)\! \frac{ \partial_i \ddot{\rho} }{D \nabla^2} + \frac{ E - J }{D} \partial_i \rho \nonumber\\
&+& \frac{D-C}{\ks} \partial_i \nabla^2 \phi \label{CETheoryA}
\eea

{\em Enforcing locality}: 
From the previous equation it should be clear that we must require $AB=IH$ in order for the sources themselves to be local. It turns out we also need to enforce $D=C$ to remove the Newtonian term from this otherwise modified conservation law. The reason for this can be seen ahead in eq.~(\ref{phiAsoln}); the Newton potential $\phi$ has a term in it $\propto\ddot{\rho}/\nabla^4$ (which is present even in the GR limit $I=K$), and would imply a non-local contribution to eq.~(\ref{CETheoryA}), unless we set $D=C$. 
Then the modified conservation equation becomes
\beq
\partial_j \tau_{ij} + \dpri = \frac{ E - J }{D} \partial_i \rho \label{consi}
\eeq	
This appears to still violate source conservation. However, as mentioned earlier, in this framework the trace of our source $\tau_{ij}$ is another scalar under rotations and is not completely distinguishable from $\rho$. Without loss of generality, we can write 
\beq
\tau_{ij}=\tilde{\tau}_{ij}+{E-J\over D}\delta_{ij}\rho
\label{tildetau}\eeq 
Note that this is a special case of eq.~(\ref{tauTF}) with $C=D$ and evidently maintains linear coupling to the graviton $h_{\mu\nu}\mathcal{T}^{\mu\nu}=\tilde{h}_{\mu\nu}\tilde{\mathcal{T}}^{\mu\nu}$ with $\tilde\phi=\phi+h(E-J)/D$. 
We then insert this into the above equation to obtain
\beq
\partial_j \tilde{\tau}_{ij} + \dpri = 0
\label{tildetauCons}\eeq	
Hence for all intents and purposes, the sources are conserved, though it is not apparent when written in terms of the original variable $\tau_{ij}$. Alternatively we can simply make the canonical definition that occurs in GR, namely $E=J$ (which in fact can be done without loss of generality, as discussed below eqs.~(\ref{tauTF},\ref{phiTF})), however, for completeness we do not impose this condition here.
Having made these choices
\beq
AB=IH ~,~~  C=D \label{firstconditions}
\eeq	
we can find the inhomogeneous solutions (ignoring external gravitons) of the equations of motion (\ref{rhoA}--\ref{pA}) with the conditions (\ref{firstconditions}) to obtain
\bea			
{h\over\ks} &=& \frac{ - \rho}{D\nabla^2} \label{hAsoln}\\
{\psi_i\over\ks} &=& \frac{p_i}{H \nabla^2} + \frac{B}{DH} \frac{\partial_i \dot{\rho}}{\nabla^4} \label{psiAsoln}\\
{\phi\over\ks} &=& \frac{-\tau}{2D \nabla^2} + \frac{3I-K}{2D^2} \frac{\ddot{\rho}}{\nabla^4} + \frac{E + L - 3J}{2D^2} \frac{\rho}{ \nabla^2} \label{phiAsoln}\\
{h_{ij}\over\ks} &=& \frac{\tau_{ij}}{\square} + \frac{\left( \partial_i \partial_j - \delta_{ij} \nabla^2 \right) \tau}{2 \square \nabla^2} + \frac{B}{I} \frac{\partial_{(i} \dot{p}_{j)}}{\square \nabla^2}\nonumber\\
&+&\frac{\delta_{ij}}{2D} \frac{\left[ \left( E+L-J \right) \nabla^2 + \left(I - K \right) \partial_t^2 \right] \rho}{\square \nabla^2} \nonumber\\
&+& \frac{\left[ \left( K + I \right) \partial_t^2 + \left( 3J - 3E -L \right) \nabla^2 \right] \partial_i \partial_j \rho}{2D \square \nabla^4} \,\,\,\,
\label{gaugeFields}
\eea
where 
\beq
\square \equiv K \partial_t^2 - L\nabla^2. 
\eeq	
We then use these solutions to eliminate the fields from the interaction Lagrangian and integrate by parts in the action to replace all divergences using the conservation equations (\ref{cons0}) and (\ref{consi}). The tree-level exchange action $-{1\over4}\kappa h_{\mu\nu}\T^{\mu\nu}$ then becomes
\bea
\frac{8\Lex}{\kappa^2} &=& \tau_{ij} \frac{\tau_{ij}}{\square} - \frac{\tau}{2} \frac{\tau}{\square} 
+ \frac{\rho}{2} \bigg[2a_5 \nabla^2 \partial_t ^2+a_6 \nabla^4 + a_7\partial_t^{4}  \bigg] \frac{ \rho }{\square \nabla^4}  \nonumber\\
&-& 2 p_i \!\left[ \frac{a_3\nabla^2 + a_4\partial_t^2}{\square \nabla^2} \right] \!p_i 
+\rho\!\left[\frac{ a_1\nabla^2 + a_2\partial_t^2 }{\square \nabla^2} \right]\!\tau
\label{nonlocal1}
\eea	
where $a_1=(E-J+L)/D$, $a_2=(I-K)/D$, $a_3=L/H$, $a_4=(-K + H I^2 / B^2)/H$, $a_5=(2JK + E(3I-K) - 3IJ + L(2I-K-2B^2/H)) /D^2$, $a_6=(2 (2J -E) L - 3(E-J)^2 - L^2)/D^2$, and $a_7=(I^2 + 4IK - K^2 - 4B^2 K /H)/D^2$.  
The first two terms are clearly local, but the other terms contain nonlocal pieces which we wish to eliminate. These terms cannot be combined further to cancel the non-localities, without placing unphysical restrictions on the sources. So the coefficients must be constrained to make interactions local. In each nonlocal term, this can be accomplished by requiring the coefficient of the $\partial_t^2$ pieces vanish, so that the Laplacians cancel, leaving only an inverse box operator. In particular, the $\rho \tau$ term requires $I=K$, then the $p_i p_i$ term requires $B^2 =H K$ (which implies $A=B$), and the $\rho\rho$ term requires $E=(J + L)/2$.
Enforcing these three new conditions
\beq
I=K ~,~~ B^2=HK ~,~~ E=(J+L)/2
\eeq
on the coefficients gives
\beq
\frac{8\Lint}{\kappa^2} = \tau_{ij} \frac{\tau_{ij}}{\square} - \frac{\tau}{2} \frac{\tau}{\square} + a_1\rho \frac{\tau}{\square} - 2a_3 \,p_i \frac{p_i}{\square} + a_6 \frac{\rho}{2} \frac{\rho}{\square}  \label{localLint}
\eeq
where $a_1=(3L-J)/(2D)$, $a_3=L/H$, and $a_6=(-3 J^2 + 18 J L - 11 L^2)/(4 D^2))$. This is now completely local and has a similar form to the GR action eq.~(\ref{Lint_GR}). We are left with five unspecified coefficients, $D,H,J,K,$ and $L$. 

This is almost of the form of (\ref{Lint_GR}), except for 3 differing prefactors: $a_1$, $a_3$, $a_6$. It can almost be put into identical form by a re-scaling of sources $\rho$ and $p_i$. However that would leave 1 residual parameter left over (we do not re-scale $\tau_{ij}$ here because the $\tau_{ij}\tau_{ij}$ and $\tau\tau$ terms are already canonical). Moreover, we said that our sources are not quite conserved. This can all be fixed by expressing the exchange action in terms of the conserved source $\tilde\tau_{ij}$ (from eqs.~(\ref{tildetau},\ref{tildetauCons})), giving
\beq
\frac{8\Lint}{\kappa^2} = \tilde{\tau}_{ij} \frac{\tilde{\tau}_{ij}}{\square} - \frac{\tilde{\tau}}{2} \frac{\tilde{\tau}}{\square} + {L\over D}\rho \frac{\tilde{\tau}}{\square} - {2L\over H} \,p_i \frac{p_i}{\square} + {L^2\over D^2}\frac{\rho}{2} \frac{\rho}{\square}  \label{localLint2}
\eeq
(or, equivalently, setting $E=J$). We can then re-scale $\rho\to\rho\,(D/L)$ and $p_i\to p_i\sqrt{H/L}$, and identifying that $\rho,\, p_i,\,\tilde{\tau}_{ij}$ obey the same conservation laws as $T^{00},\,T^{0i},\,T^{ij}$, respectively, we obtain the GR action in its exact form. Furthermore, when using $\tilde\tau_{ij}$ (or equivalently, setting $E=J$), we have $\partial_\mu\T^{\mu\nu}=0$. We therefore recover GR exactly to this order of analysis.

\subsection*{Including Mass Terms}

So far we have not included mass terms in our starting action eq.~(\ref{General}). However, rotation invariance allows one to include 5 different types of mass terms
\begin{equation}
\mathcal{L}_{m}=-{1\over2}m_1^2\phi^2-m_2^2 \phi h-{1\over2}m_3^2 h^2-{1\over2}m_4^2\psi_i\psi_i-{1\over2}m_5^2h_{ij}h_{ij}
\end{equation}
Since we are projecting down to 2 degrees of freedom and not assuming Lorentz invariance, then a priori, all 5 are in fact allowed. The analogous terms were included in our Part 1 paper (Ref.~\cite{Accompanying}) on electromagnetism, although in that case only 2 terms are allowed. 

However, just like in the electromagnetic case, they all lead to non-locality. One can see this as follows: We are interested in deforming away from GR in which all the masses are zero. So let's consider the situation in which the masses are small. We can then begin by operating in a regime of length scales $L\ll 1/m$, so that we can be sure the corrections from masses are irrelevant and the above constraints from locality still apply. This leads to the usual conservation laws, as we showed above (with the linear shift on $\tau_{ij}$ if $E\neq J$). By then including finite corrections from the masses, the conservation laws now become
\begin{eqnarray}
\partial_i p_i+\dot{\rho}_r &=& (m_1^2\dot{\phi}+m_2^2\dot{h})(A/D)\\
\partial_j\tilde{\tau}_{ij}+\dot{p}_{r,i}&=&m_2^2\partial_i\phi+m_3^2\partial_ih+m_4^2\dot{\psi}_i(A/H)
\end{eqnarray}
At leading order in the masses, we know that $\phi,\,\psi_i,\,h$ are all non-local (even in the GR limit); see eqs.~(\ref{hAsoln}-\ref{phiAsoln}). This means that if we insert this into the above pair of continuity equations, the right hand sides will be non-local if any of the $m_1,\ldots,m_4$ are non-zero. This means the sources are non-local and hence the theory is non-local, unless
\begin{equation}
m_1=m_2=m_3=m_4=0
\end{equation}

Our only remaining consideration then is $\mathcal{L}_{m}=-{1\over2}m_5^2h_{ij}h_{ij}$. To understand its consequences, we can just note that it has a similar structure to the already present term in the action $\Delta\mathcal{L}=-{1\over2}L\,\partial_k h_{ij}\partial_k h_{ij}={1\over2}h_{ij}(L\nabla^2)h_{ij}$ (plus boundary term). Hence the consequences of this mass term are equivalent to the replacement:
\begin{equation}
L\to L-{m_5^2\over\nabla^2}
\end{equation}
in the existing results. By making this replacement in eq.(\ref{localLint2}) we are immediately led to non-local terms due to the inverse Laplacian. Hence we also need
\begin{equation}
m_5=0
\end{equation}
along with all the other masses vanishing too, as described above. A similar analysis applies to mass terms in the other upcoming theories too, but we suppress the details for simplicity and will ignore the masses for the remainder of the paper.

\section{Theory B: Constraint from equations of motion}\label{grcase2}

We now wish to explicitly allow non-conservation of $\T^{\mu\nu}$. We do this by using the full equations of motion (\ref{equations of motion1}--\ref{equations of motion}) before gauge-fixing to again write conservation equations of the form of  eqs.\,(\ref{cons0}) and (\ref{consi}) which we explicitly allow be nonzero functions, $\sigma$ and $w_i$. In this theory $\partial_i \psi_i$ and $\partial_j h_{ij}$ will be fixed in terms of $\sigma$ and $w_i$. We can then solve the equations of motion in this theory to write the interaction Lagrangian just in terms of sources, and find for general $\sigma$ and $w_i$ the only way this theory can be local is if $\sigma$ and $w_i$ vanish, recovering conservation of $\T^{\mu\nu}$. In that case we again uniquely recover GR by enforcing locality. However, if we allow $\sigma$ and $w_i$ themselves to be derivatives of some local functions, we find locality requires the theory reduce to GR with some additional terms.
	
Returning to the theory with all 12 unknown coefficients, we use the general equation for the non-conserved scalar source eq.~(\ref{sigmaEq}) and impose $BD=AC$ in order to be able to use this to fix $\partial_i \psi_i$ to
\beq
\partial_i \psi_i = \frac{\ks \, \sigma}{(H-F)\nabla^2} \label{divpsi}
\eeq
In the next section, we will generalize this condition (see ahead to eq.~(\ref{dpsiC})), but for the sake of clarity, we will make this simplification here as it will not affect our qualitative results.
Similarly, we use the general equation for the non-conserved vector source eq.~(\ref{wEq}) and impose $C=D$ and $I=K=A^2/H$ in order to fix $\partial_j h_{ij}$ to
\bea
\partial_j h_{ij} 
&=& \frac{\ks}{(G-L) \nabla^2} \Big{[}\frac{2E-J-G}{2G+J-2E-L} \frac{\partial_i q}{\nabla^2} - \frac{A}{H} \frac{\partial_i \dot{\sigma}}{\nabla^2} \nonumber\\
&&\,\,\,\,\,\,\,\,\,\,\,\,\,\,\,\,\,\,\,\,\,\,\,\,\,\,\,\,\,\,\,+\frac{J-E}{D} \partial_i \rho + w_i \Big{]} \label{divhij}
\eea
where for convenience we have defined 
\beq
q \equiv \partial_j w_j - \frac{A}{H} \dot{\sigma} + \frac{J-E}{D} \nabla^2 \rho \label{qdef}
\eeq
By requiring that the equations of motion fix $\partial_i \psi_i$ and $\partial_j h_{ij}$, and hence cut down to 2 degrees of freedom, the number of unknown coefficients has been reduced by enforcing
\beq
A=B,~~~ C = D,~~~ I = K = A^2 / H
\eeq
This means the number of parameters is down to $12-4-1-4=3$ in the (classical) theory. On the other hand, the sources are described by 2 arbitrary functions $\sigma$ and $w_i$; so there is considerable freedom in the theory.

{\em Enforcing locality}: 
Similar to the previous section, we solve the equations of motion to obtain the inhomogeneous solutions for $\phi$, $\psi_i$ and $h_{ij}$. These results are somewhat complicated and are reported in Appendix \ref{A}. We then use these to write the exchange action only in terms of the sources, integrating by parts to replace divergences using the definitions of $q$, $\sigma$, and $w_i$. Doing this, the tree-level exchange action may be written in terms of the sources $\T^{\mu\nu}$ (built out of $\rho, p_i, \tau_{ij}$) and the non-conservation parameters $\sigma$ and $w_i$. We find it has the form
\bea
\frac{8\Lex}{\kappa^2} &=& \tau_{ij} \frac{ \tau_{ij}}{\square} - \frac{\tau}{2} \frac{ \tau }{\square} + {L \over D} \rho {\tau \over \square} - \frac{2L}{H} p_i \frac{p_i}{\square} + {4A \over H} p_i {\dot{w}_i \over \square \nabla^2}  \nonumber\\
&+& {\rho \over 2} \left[ {b_1 \nabla^2 + b_2 \partial_t^2 \over \square \nabla^2} \right] \rho  + {2 w_i \over G-L } \left[ {G \nabla^2- K \partial_t^2 \over \square \nabla^4} \right] w_i \nonumber\\
&+& \! \sigma \left[ {b_3 \nabla^4 + b_4 \nabla^2 \partial_t^2 + b_5 \partial_t^4 \over \square \nabla^6} \right] \sigma + \! \rho \! \left[ {b_6 \nabla^2 + b_7 \partial_t^2 \over \square \nabla^4} \right] \! \dot{\sigma} \nonumber\\
&+& \! \partial_i w_i \! \left[ {b_8 \nabla^2 + b_9 \partial_t^2 \over \square \nabla^6} \right] \! \partial_j w_j + \! \rho \left[ {b_{10} \nabla^2 + b_{11} \partial_t^2 \over \square \nabla^4} \right] \! \partial_i w_i \nonumber\\
&-& {A \over H} \tau {\dot{\sigma} \over \square \nabla^2} + \tau { \partial_i w_i \over \square \nabla^2} + \dot{\sigma} \left[ { b_{12} \nabla^2 + b_{13} \partial_t^2 \over \square \nabla^6 } \right]\! \partial_i w_i
\label{LexB}\eea
where the coefficients $b_1,b_2,\ldots b_{13}$ are given in Appendix \ref{A}.
The first 4 terms have the same form as eq.~(\ref{Lint_GR}) and are clearly local. 

The $w_i w_i$ and $\sigma\,\sigma$ terms involve inverse Laplacians. If the functions $\sigma$ and $w_i$ are general functions, then the theory is immediately non-local. The most direct way to avoid this problem is to impose that they vanish, i.e., $\sigma=w_i=0$. 
This means the theory readily reduces to GR as it now becomes similar in structure to the previous Theory (A). It only requires one additional constraint on parameters to remove the non-local part of the $\rho\,\rho$ term.

However, if we suppose that $\sigma$ and $w_i$ are not general functions, but are instead given in terms of spatial derivatives of other local functions, then there is a possibility to cancel the inverse Laplacians and maintain locality. We find that the necessary condition to obtain a local action is that $\sigma$ and $w_i$ can be expressed in terms of local functions $f$ and $g_i$ as follows
\bea
\sigma&=&(F-H)\nabla^2f\\
w_i&=&(G-L)(\nabla^2g_i+\partial_i\partial_jg_j)+{A\over H}(F-H)\partial_i\dot{f}
\eea
where the prefactors, $(F-H)$ and $(G-L)$ are for convenience (as expanded on later), but could be re-absorbed into $f$ and $g_i$ if desired. By inserting this into the exchange action (\ref{LexB}), we find that the action becomes local with just one more condition required to eliminate the non-local $\rho \, \rho$ term, namely $b_2 = 0$.
We choose $E=J$ for simplicity of presentation, though it is not required, and then the action simplifies into the following form
\bea
\frac{8\Lint}{\kappa^2} &=& \tilde\tau_{ij} \frac{\tilde\tau_{ij}}{\square} - \frac{\tilde\tau}{2} \frac{\tilde\tau}{\square} + {L\over D}\tilde\rho \frac{\tilde\tau}{\square} - {2L\over H} \,\tilde{p}_i \frac{\tilde{p}_i}{\square} + {L^2\over D^2} \frac{\tilde\rho}{2} \frac{\tilde\rho}{\square}  \label{localLint}\nonumber\\
&+&2(F-H)f^2-2(G-L)g_ig_i
\label{LintB}\eea
where the sources with the tilde overbar indicate that they are conserved in the usual sense, i.e., if we form $\tilde\T^{\mu\nu}$ out of them, then we have $\partial_\mu\tilde\T^{\mu\nu}=0$. They are related to our original sources by
\bea
\tau_{ij}&=&\tilde\tau_{ij}
+(G-L)\partial_{(i} g_{j)} \label{tauTB}\\
p_i&=&\tilde{p}_i+(F-H)\partial_i f\label{pTB}\\
\rho&=&\tilde{\rho}\label{rhoTB}
\eea
We note that if we set $f=g_i=0$, this recovers exactly the result of the previous section in eq.~(\ref{localLint2}), which we already remarked is equivalent to GR under a re-scaling of sources.

On the other hand, for non-zero $f$ and/or $g_i$ our result for the exchange action in eq.~(\ref{LintB}) clearly differs from the result in GR due to the presence of these new \emph{ultra}-local terms on the 2nd line, which have no GR analogue. We shall return to discuss these terms in Section \ref{Discussion}, where we will explain how these are in fact consequences of purely decoupled sectors, and do not actually represent a meaningful modification of GR.

\section{Theory C: Generalized Constraint}\label{grcase3}
		
There exists a third distinct option to cut down to 2 degrees of freedom by combining the approaches of the previous two sections. In Sec.\,(\ref{grcase1}) we set $\partial_i \psi_i = \partial_j h_{ij} = 0$. One might wonder whether there is a way to similarly ``gauge-fix'' the fields without forcing the divergences to vanish. However, if we arbitrarily declare $\partial_i \psi_i = \mathcal{F}_0$ and $\partial_j h_{ij} = \mathcal{F}_i$ where $\mathcal{F}_\mu$ are some functions of the matter fields $\Psi_m$, this will not in general be consistent with the equations of motion which give something of the form of eqs.\,(\ref{divpsi}) and (\ref{divhij}) (or more general if fewer constraints on the coefficients $A,\mathellipsis,L$). In this section we find a general form of $\mathcal{F}_\mu = \mathcal{F}_\mu [\Psi_m]$ consistent with the equations of motion and use this to cut down to 2 degrees of freedom of the graviton. 
	
Returning to the full equations of motion (\ref{equations of motion}), without any conditions on the coefficients we can write
\beq
\partial_i\psi_i = {\ks\,\sigma\over (H-F)\nabla^2}+{(BD-AC)\over D(H-F)\nabla^2}\!\left[{\ks\,\dot\rho\over C} - \partial_i \partial_j \dot{h}_{ij} \right]
\eeq
which immediately fixes $\partial_i\psi_i$ in terms of the sources and $\partial_j h_{ij}$. So it only remains to fix $\partial_j h_{ij}$. Similarly to the previous sections we can obtain a general expression for $w_i$ that parameterizes vector source violation. 
As before, we must set $D=C$ to eliminate the Newtonian term. We can then solve for $\partial_j h_{ij}$ as
\beq
\partial_j h_{ij} = \M_{ij} \left[ -\epsilon {\partial_j \dot{\sigma} \over \nabla^2} -\alpha {\partial_j \ddot{\rho} \over \nabla^2} - \lambda {\partial_j \rho\over D} + w_j \right] \label{divh3}
\eeq	
where the matrix valued differential operator $\mathcal{M}_{ij}$ is
\beq
\M_{ij} 
 \equiv {1\over \square_1\! } \left( \delta_{ij} - { \square_2 \over \square_1 + \square_2 } {\partial_i \partial_j \over \nabla^2 }\right) 
\eeq
with a pair of wave-like operators, defined as
\beq		 
\square_1 \equiv \gamma \partial_t^2 + \delta \nabla^2,~~~~ \square_2 \equiv \alpha \partial_t^2 + \beta \nabla^2
\eeq
In the above set of equations we have defined some convenient collections of the coefficients $A,\mathellipsis,L$
\bea
\alpha &\equiv & {2(A-B)^2\over F-H}+{A^2\over H}-I,\,\,\,\, 
\delta \equiv G-L,\,\,\,\,\gamma \equiv K- {A^2 \over H},\nonumber\\
\beta &\equiv& J+G-2E,\,\,\,\,\epsilon \equiv {2(A-B)\over F-H}+{A\over H},\,\,\,\,\lambda \equiv E-J\,\,\,\,\,\,\,\,\,\,\,\,
\eea
Now the issue is that, despite appearances, $\partial_j h_{ij}$ is still dynamical, since it is given in terms of an inverse wave-like operator contained in the denominator of the definition of $\M_{ij}$.  In order to make $\partial_j h_{ij}$ actually non-dynamical and thereby cut down to 2 degrees of freedom, we need to either (i) make $\square_1$ not involve time derivatives, i.e., set $\gamma=0$. However this would simply return us to the basic structure of Theory (B) of the last section. 

So instead we need to (ii) make the $\square_1$ wave-like operator cancel out. For the $\sigma$ and $w_i$ terms in eq.~(\ref{divh3}), this will occur if they are chosen to be proportional to $\square_1+\square_2$, which we parameterize as follows
\bea
\epsilon \, \sigma &= & \left( \square_1 + \square _2 \right) \f \label{fdef}\\
w_j &= &  \left( \square_1 + \square _2 \right) {\partial_j \g \over \nabla^2} \label{gdef}
\eea
where $s_1$ and $s_2$ are scalars. Note that here
we have needed to enforce that the $w_i$ term is proportional to the gradient of a scalar $\g$ in order for the cancellation to occur. This leaves only the $\rho$ terms in eq.~(\ref{divh3}) as a possible source that would generically make $\partial_j h_{ij}$ dynamical. Since $\rho$ is a physical source, imposing any conditions on it would over-constrain the theory, so the only option is to set 
\beq
-\alpha \partial_t^2 + \lambda \nabla^2 = \om \left[ \left( \alpha + \gamma \right) \partial_t^2 + \left( \beta + \delta \right) \nabla^2 \right]  
\eeq
where $\om$ is a constant of proportionality (this means $-\alpha=\om(\alpha+\gamma)$ and $\lambda=\om(\beta+\delta)$). This procedure has now completely fixed $\partial_j h_{ij}$ and $\partial_i\psi_i$ to the following non-dynamical values
\bea	
\partial_j h_{ij} &=& \ks{\partial_i \over \nabla^2 } \left( \g-\dot{s}_1 - {\omega \rho \over D} \right)\label{hijEq} \\
\partial_i \psi_i &=&\ks \frac{ \left( \square_1 + \square_2 \right) \f}{\epsilon \left( H-F \right) \nabla^2} +\ks\, \zeta{\left[\dot{s}_2-\ddot{s}_1 - {\dot{\rho} \over D}(\omega+1) \right] \over \nabla^2}\label{dpsiC}\,\,\,\,\,\,\,\,
\eea
where $\zeta\equiv(A-B)/(H-F)$.

{\em Enforcing locality}: 	
As in the previous sections, we now rewrite the exchange action just in terms of the sources, with their non-conservation parametrized now by $\f$ and $\g$ as defined by eqs.~(\ref{fdef}) and (\ref{gdef}). The result contains many nonlocal terms, whose structure is sketched in Appendix \ref{B}. The necessary conditions for the nonlocal terms to vanish are $\alpha=0$, $\gamma=0$ and a condition relating $J$ to $L,G,$ and $\omega$. We can summarize these conditions as
\bea
K&=& {A^2\over H},\,\,\,\,\,I=K+{2(A-B)^2\over(F-H)},\nonumber\\
J&=&{L(1+2\omega-\omega^2)+2G\omega^2\over(1+\omega)^2}
\eea

However we find that for generic $\f$ and $\g$ it is impossible to obtain locality. This is not surprising, given the form of eq.~(\ref{hijEq}) in which $\partial_j h_{ij}$ would be non-local itself (even if $\omega=0$). Therefore we require
\beq
\g=\dot{s}_1 +\nabla^2s_3
\eeq
where $s_3$ is some local scalar function. When inserted into the action, we find that everything is now local. The sources $p_i$ and $\tau_{ij}$ are once again not directly conserved, due to $\sigma$ and $w_i$ being non-zero. Nevertheless, similar to the previous Theory (B), we can readily relate them to conserved sources $\tilde{\tau}_{ij},\,\tilde{p}_i$, as
\bea
\tau_{ij} &=& \tilde{\tau}_{ij}+(\delta+\beta)\,\delta_{ij}\!\left[\chi\,\dot{s}_1+\partial_i\partial_j s_3\right]+\mathcal{P}_{ij}[s_3]\label{tauC}\\
p_i &=& \tilde{p}_i+{\beta+\delta\over\epsilon}\,\partial_is_1\label{pC}\\
\rho&=&\tilde{\rho}
\eea
where $\mathcal{P}_{ij}[s_3]\equiv\omega(\delta+\beta)\left[\partial_i\partial_j-\delta_{ij}\nabla^2\right]\!s_3$ is an identically conserved quantity, which is useful to fully diagonalize the system, and $\chi\equiv1-A/(H\epsilon)$ (note that if $A=B$, then $\chi=0$).

The final result for the tree-level exchange action is then found to be exactly of the familiar GR terms, plus a pair of ultra-local terms. For simplicity, we mention the $A=B$ and $E=J$ form of the ultra-local terms, which are
\beq
{8\Delta\Lex\over\kappa^2} = {8H^2(G-L)^2\over A^2(F-H)}\,s_1^2-2(G-L)(\partial_is_3)^2
\eeq
In fact this is related to the result of Theory (B) in eq.~(\ref{LintB}), with the identifications
\beq
f ={2H(G-L)\over A(F-H)}\, s_1,\,\,\,\,\,\, g_i=\partial_i s_3
\eeq
So again we almost recover GR, except for a pair of additional terms, which we discuss in Section \ref{Discussion}.

\section{Theory D: Transverse-Traceless}\label{TT}

In this section, for completeness, we will discuss the one final term that we could have added to the generalized action (\ref{General}) that is compatible with rotation invariance, namely
\beq
\Delta\Lag = -{1\over 2}M(\partial_i\phi)^2
\label{NewTerm}\eeq
This term is not present in the GR action and hence we did not study it previously in this paper. But for the sake of completeness, let us examine this briefly now. 

Although one could perform a more general analysis, we will use this extra term to focus on a qualitatively new way of cutting down to 2 degrees of freedom. We will impose the transverse-traceless ``gauge" choice
\beq
\partial_i h_{ij}=0,\,\,\,\,\,\,h=0
\eeq
In fact these choices are in some sense the most natural way to cut down to 2 degrees of freedom starting with the symmetric polarization matrix $h_{ij}$, while the fields $\phi$ and $\psi_i$ are non-dynamical. As is well known, in GR the transverse-traceless gauge is not a gauge that is allowed in general as it must be violated inside of matter. However, by deforming away from GR with the term in eq.~(\ref{NewTerm}), it now becomes possible to implement this gauge fixing both inside and outside of matter, as we will explore here. This makes this final choice special, because by imposing the transverse-traceless conditions everywhere, this theory cannot recover GR in any non-trivial limit. 

Having imposed the transverse-traceless constraint, we can solve for the fields in complete generality without needing to restrict any of the 13 parameters $A,B,\ldots,L,M$. However, the parameters $E,G,I,J$ will not appear in the classical equations of motion; this leaves us with $13-4-4=5$ parameters that affect interactions at tree-level.

The solutions are readily found to be
\bea
{\phi\over\ks} &=& - {\rho\over M\nabla^2}\\
{\psi_i\over\ks} &=&\left[\delta_{ij}+\left(F\over H-F\right){\partial_i\partial_j\over\nabla^2}\right]\!{p_j\over H\nabla^2}\\
{h_{ij}\over\ks} &=& {1\over\square}\Bigg{(}\tau_{ij}-\left[C\delta_{ij}-D{\partial_i\partial_j\over\nabla^2}\right]\!{\rho\over M}
+{A\over H}{\partial_{(i}\dot{p}_{j)}\over\nabla^2}\nonumber\\
&&-\left({2\over H-F}\right)\left[B\delta_{ij}-{AF\over H}{\partial_i\partial_j\over\nabla^2}\right]\!{\partial_k\dot{p}_k\over\nabla^2}
\Bigg{)}\,\,\,\,
\eea
By taking the trace and divergence of this final expression for $h_{ij}$, and demanding that it is transverse-traceless, we obtain the pair of equations for the sources
\bea
0&=&\tau+{(2A-6B)\over(H-F)\nabla^2}\partial_k\dot{p}_k+\left(D-3C\over M\right)\!\rho\label{ContD1}\\
0&=&\partial_i\tau_{ij}+\dpri+\left(D-C\over M\right)\!\partial_i\rho\nonumber\\
&+&\left({A(H+F)-2BH\over H(H-F)}\right)\!{\partial_i\partial_k\dot{p}_k\over\nabla^2}\label{ContD2}
\eea
For locality, we need the sources to obey local continuity type equations. In this work, we will not impose overly constraining conditions on $\partial_kp_k$, and hence we need the inverse Laplacian terms in eqs.~(\ref{ContD1},\ref{ContD2}) to vanish, so
\beq
A=3B,\,\,\,\,\,\,F=-H/3
\eeq
(which again shows this is disconnected from GR where $A=B$ and $F=H$).

With these conditions, we can then form the tree-level exchange action. There are a number of terms, but for simplicity, we here report on only the terms that are proportional to $(\partial_ip_i)^2$; these are found to be
\beq
{8\Lag_{(\partial_ip_i)^2}\over\kappa^2}=-{\partial_ip_i\over 2H}\!\left[{A^2\over H}\partial_t^2-K\partial_t^2+L\nabla^2\right]\!
{\partial_j p_j\over\square\nabla^4}
\eeq
This is clearly non-local and so it must vanish to avoid instantaneous action at a distance. We can make the piece $\propto \partial_t^2/(\square\nabla^4)$ vanish, by setting $K=A^2/H$. However, to make the piece $\propto 1/(\square\nabla^2)$ vanish, we would require
\beq
L=0
\eeq
This means the graviton speed would have to vanish. This is an extreme way to build a local theory, by preventing any finite speed propagation altogether. Such a theory is of little interest and we do not pursue it further. Hence we conclude that our starting point with the new term that deviates from GR in eq.~(\ref{NewTerm}) is unacceptable.

\section{Discussion}\label{Discussion}

In this work we have imposed locality on theories involving spin 2 particles (gravitons), without assuming Lorentz boost symmetry. In Theories (A), (B), and (C) we have recovered the form of the leading tree-level exchange action of GR, although in both Theories (B) and (C), there were additional terms (while Theory (D) was a trivial theory in the end).

\subsection*{Additional Terms}

In the most general version of Theory (B), we found we could have an arbitrary scalar function $f$ and an arbitrary vector function $g_i$ which parameterize different ways of violating source conservation; see eqs.~(\ref{tauTB},\ref{pTB}) (and in Theory (C) we can have arbitrary scalar functions $s_1$ and $s_3$; see eqs.~(\ref{tauC},\ref{pC}), while no such terms were allowed in Theory (A)). Such additional terms are quite analogous to the additional term that arises in our accompanying paper on electromagnetism \cite{Accompanying}. We can understand them in a similar fashion as follows.

Firstly, let us return to the regular GR action for the graviton, plus conserved sources $\partial_\mu\tilde{\T}^{\mu\nu}=0$, and a pair of additional terms, as follows
\beq
\Lag =
\Lkin-{\kappa\over2}h_{\mu\nu}\tilde\T^{\mu\nu}+\ks^2(F-1)f^2-\ks^2(G-1)\,g_i g_i
\label{GRUL}\eeq
These additional terms are evidently completely decoupled sectors, expressed in terms of functions $f$ and $g_i$, whose prefactors are for convenience. 
Since the regular GR action exhibits gauge invariance, we can make any gauge choice we desire. To illustrate the connection to our earlier theories, it is useful to make the following gauge choices
\beq
\partial_i\psi_i = -\ks\,f,\,\,\,\,\,\partial_j h_{ij}=\ks\, g_j
\eeq
We can use these conditions to construct the identity $\ks^2f^2=-(\partial_i\psi_i)^2-2\ks\,f\partial_i\psi_i$, as well as a similar identity for $\ks g_ig_i$, and then the action can be written (after an integration by parts on the second term)
\beq
\Lag=\Lkin-{\kappa\over2}h_{\mu\nu}\T^{\mu\nu}-(F-1)(\partial_i\psi)^2+(G-1)\partial_j h_{ij}\partial_k h_{ik}
\label{GRRW}\eeq
where the source is identified as
\beq
\T^{\mu\nu}=\tilde{\T}^{\mu\nu}+\delta\T^{\mu\nu}
\eeq
where $\delta{\T}^{\mu\nu}$ are additional non-conserved pieces, precisely those of the form identified earlier in Theory (B) \& (C) in eqs.~(\ref{tauTB},\ref{pTB}) \& (\ref{tauC},\ref{pC}) (here we are taking the special case $E=J=H=L=1$ for simplicity of presentation), i.e., $\delta\T^{ij}=(G-1)\partial_{(i} g_{j)}$, $\delta\T^{0i}=(F-1)\partial_i f$, and $\delta\T^{00}=0$. We note that this identity is useful because it allows us to break up the decoupled sectors into 2 pieces: one piece that goes into the final terms of eq.~(\ref{GRRW}) and another piece that goes into a shift in $\mathcal{T}^{\mu\nu}$. 
With this identification the action in eq.~(\ref{GRRW}) becomes precisely a re-writing of Theory (B) and (C), in the special case: $A=B=C=D=E=H=I=J=K=L=1$, general $F,\,G$, and non-conserved sources provided by arbitrary $f$ ($\propto s_1$) and $g_i$ ($\propto\partial_i s_3$). Since our starting point to construct this was manifestly local in eq.~(\ref{GRUL}), it is obvious from this point of view that Theory (B) should allow for this local construction. This provides a non-perturbative proof that the contributions from $f$ and $g_i$, which were seen to decouple at the level of the tree-level exchange action in eq.~(\ref{LintB}), in fact persists as an exact statement, because the starting action (\ref{GRUL}) shows they are completely decoupled sectors. In this sense, these ``corrections" to GR that appeared in Theory (B), and related ``corrections" that appeared in Theory (C), do not constitute physical modifications at all.

\subsection*{ Lorentz Symmetry}
In this work, we did not a priori assume anything about the structure of the sources $\T^{\mu\nu}$. However by imposing the most basic notion of locality, that we do not have instantaneous action at a distance when coupling to a spin 2 particle (graviton), we have shown that a necessary condition is that it is conserved $\partial_\mu\tilde{\T}^{\mu\nu}=0$ (we only need to comment on $\tilde{\T}^{\mu\nu}$ here, rather than the full $\T^{\mu\nu}$, as the differences are only associated with re-definitions and/or irrelevant decoupled sectors, as discussed above). 

Now one can explore the ramifications of needing the sources to exhibit local conservation of this variety. Firstly, we have assumed in this work that the laws of physics exhibit translation invariance (see more about relaxing that assumption below). As is well known, this implies the conservation of the energy-momentum tensor by the Noether theorem. However, it is important to emphasize that by itself this only means there is an object with {\em mixed} indices that is conserved, i.e.,
\beq
\partial_\mu T^{\mu}_{\,\,\nu}=0
\eeq
So naturally there are 4 conserved currents, labelled with index $\nu$ here. And so there are 4 conserved quantities, which are the familiar total energy and total momentum 
\beq
E=\int\! d^3x\,T^{0}_{\,\,0},\,\,\,\,\,\,
P_i=\int\! d^3x\,T^{0}_{\,\,i}
\eeq 
Such quantities do not rely on the existence of Lorentz invariance and so they exist even in non-relativistic condensed matter systems involving fluctuations around a translationally invariant medium. In ordinary circumstances this mixed index energy-momentum tensor cannot be lifted to any symmetric object. For example, consider the following theory of 2 coupled scalars
\begin{equation}
\mathcal{L}_\varphi=\sum_{n=1}^2\left({1\over2}\dot{\varphi}_n^2 -{1\over2}c_n^2(\nabla\varphi_n)^2\right)-\lambda\,\varphi_1^2\varphi_2^2
\end{equation}
The theory is non-Lorentz invariant if $c_1\neq c_2$. It does have translation invariance and so it has a conserved energy-momentum tensor (really, just a matrix)
\bea
&&T^0_{\,\,\,0}=\sum_{n=1}^2\dot{\varphi}_n^2-\mathcal{L}_\varphi,\,\,T^i_{\,\,\,j}=-\sum_{n=1}^2c_n^2\partial_i\varphi_n\partial_j\varphi_n-\delta^i_j\mathcal{L}_\varphi\,\,\,\,\,\,\,\,\,\\
&&T^0_{\,\,\,i}=\sum_{n=1}^2\dot{\varphi}_n\partial_i\varphi_n,\,\,T^i_{\,\,\,0}=-\sum_{n=1}^2c_n^2\dot{\varphi}_n\partial_i\varphi_n
\eea
There is no way to make this symmetric and conserved on both indices. Note that $T^0_{\,\,\,i}$ is not proportional to $T^i_{\,\,\,0}$, as it would be in the Lorentz invariant case when all the $c_n$ are equal and can be factorized. Put differently, there is no universal Minkowski metric inverse $\eta^{\mu\nu}$ that one can use to raise the $\nu$ index and build a symmetric and conserved tensor. 

However, what we have identified in this work is that in order to preserve a primitive form of locality, the graviton (associated with a symmetric $h_{\mu\nu}$) must couple to a conserved {\em symmetric} object $\tilde{\T}^{\mu\nu}$. In order for such an object to even exist in a non-trivial theory we therefore need more than just translation symmetry. We in fact need an additional symmetry, which is that of boost invariance \cite{Jackiw}. Related details were laid out by us earlier in Ref.~\cite{Hertzberg:2017nzl}, but we can briefly illustrate this point here. From the asymptotic past to the future, translation symmetry ensures the mixed energy-momentum tensor of classical point particles
\beq
T^{\mu}_{\,\,\nu}({\bf x},t)= \sum_n v^\mu_n\, p_{n,\nu}\,\delta^3({\bf x}-{\bf x}_n)
\eeq
is conserved, with $v_n^\mu\equiv(1,{\bf v}_n)$ and $p_{n,\nu}\equiv(E_n,{\bf p}_n$) (we emphasize that so far this does not rely on Lorentz symmetry). Now in order to build a {\em symmetric} conserved quantity, we must be able to ``push" the $\nu$ index upstairs. For this to produce a symmetric object, and hence be conserved on both indices, we need
\beq
{\bf v}_n={\partial E_n\over\partial{\bf p}_n}\propto{{\bf p}_n\over E_n}
\eeq
where in the first equality we have just used Hamilton's equation and in the second step we have specified the necessary condition for the symmetric conserved tensor to exist, where the proportionality constant must be universal for all particles. The general solution of this differential equation can be put in the form $E_n^2=p_n^2c^2+m_n^2c^4$, where $c$ is a universal constant and $m_n$ (the mass) arises as an allowed constant of integration. Hence we have arrived at the dispersion relation required for Lorentz symmetry, and the full theory is indeed Lorentz invariant at this order we are working. We can then identify $\tilde{\T}^{\mu\nu}=T^{\mu\nu}$, as it is the only conserved symmetric 2-index tensor. This recovers GR at this leading order.

Furthermore, the existence of now a {\em symmetric} 2 index conserved current allows for the existence of more conserved quantities. In particular, one can now build a 3 index current:
\begin{equation}
\Theta^{\mu\nu,\lambda}\equiv x^\mu T^{\nu\lambda}-x^\nu T^{\mu\lambda}
\end{equation}
By taking its divergence, and using $\partial_\lambda T^{\mu\lambda}=0$, this is conserved
\begin{equation}
\partial_\lambda \Theta^{\mu\nu,\lambda}=0
\end{equation}
if and only if $T^{\mu\lambda}$ is symmetric.  Hence there are more conserved quantities, namely
\begin{equation}
L^{\mu\nu}\equiv \int d^3x\,\Theta^{\mu\nu,0}
\end{equation}
This is the familiar angular momentum tensor of Lorentz invariant theories. Since $L^{\mu\nu}=-L^{\nu\mu}$ is anti-symmetric, it is made out of 6 conserved quantities. 3 are the usual angular momentum (which follow trivially from our original assumptions of rotations in a preferred frame), but there are 3 more: these are the 3 conserved quantities associated with Lorentz boosts. Alternatively, by a kind of reverse Noether theorem, these 3 new conserved quantities generate the 3 Lorentz boost symmetries.

\subsection*{Future Directions}
In future work, we are interested in extending our analysis by relaxing other space-time symmetries, including time-translation symmetry. This may have potential applications to cosmology, including claimed modifications of GR that may address the cosmological horizon problem \cite{Moffat:1992ud}. Other interesting questions include the consequences of more degrees of freedom in the analysis, and exploring to what extent strong coupling problems in the UV occur, as they do in the known Lorentz invariant case.

\vspace{0.5cm}
	
\section*{Acknowledgments}
We thank Itamar Allali, Mark Gonzalez, Andi Gray, Mudit Jain, Fabrizio Rompineve, Neil Shah, and Shao-Jiang Wang for discussion.
MPH is supported in part by National Science Foundation grant No.~PHY-1720332. 

\appendix
	
\begin{widetext}
	
\section{Supplementary Details for Theory B} \label{A}
		
After solving for the inhomogeneous solutions of the equations of motion (\ref{equations of motion}), with the conditions of eqs.~(\ref{divpsi}) and (\ref{divhij}), we find
\bea
\frac{\phi}{\ks} &=& \frac{- \tau}{2D \nabla^2} + \frac{E-3J+L}{2D^2} \frac{\rho}{\nabla^2} + 
\frac{2A^2\partial_t^2+H(4E-2G-3J+L)\nabla^2}{2DH(2E-2G-J+L)} \frac{q}{\nabla^6} 
- \frac{2A}{D(F-H)} \frac{\dot{\sigma}}{\nabla^4} + \frac{A^2}{D^2 H} \frac{\ddot{\rho}}{\nabla^4}\\
\frac{\psi_i}{\ks} &=& \frac{p_i}{H \nabla^2} + \frac{A(G+J-E-L)}{DH(G-L)} \frac{ \partial_i \dot{\rho}}{\nabla^4} + \frac{F}{H(H-F)} \frac{ \partial_i \sigma}{\nabla^4} + \frac{A}{H(G-L)} \frac{\dot{w}_i}{\nabla^4} 
+ \frac{A}{H(L-G)} \frac{\partial_i \dot{q}}{\nabla^6} + \frac{A^2}{H^2 (L-G)}  \frac{ \partial_i \ddot{\sigma}}{\nabla^6}\,\,\,\,\,\,\\
\frac{h_{ij}}{\ks} &=& \frac{\tau_{ij}}{\square} - \frac{\delta_{ij}}{2} \frac{\tau}{\square} + \frac{ \partial_i \partial_j \tau}{2 \square \nabla^2}  + \frac{A}{H} \frac{\partial_{(i} \dot{p}_{j)}}{\square \nabla^2} + \frac{E-J+L}{D} \frac{\delta_{ij}}{2} \frac{\rho}{\square}+ \frac{ L(L-3J) - G(J+L) + E(G+3L) }{2(G-L)} \frac{\partial_i \partial_j \rho}{\square \nabla^2} \nonumber\\
&+& \frac{ K (2J+G-2E-L)}{2(G-L)} \frac{\partial_ i \partial_j \ddot{\rho}}{\square \nabla^4} 
- \frac{2A}{H(G-L)} \frac{ \partial_i \partial_j\dot\sigma}{\nabla^6}  + \frac{ K \partial_t^2 - G \nabla^2}{G-L} \frac{ \partial_{(i} w_{j)} }{ \square \nabla^4} + \frac{ \delta_{ij}}{2} \frac{q}{\square \nabla^2}  \nonumber\\
&+& \frac{ G(2G-2E+J) + 3(G-2E+J)L - L^2}{2(G-L)(2G-2E+J-L)} \frac{\partial_i \partial_j q}{\square \nabla^4} + \frac{K(4E-3G-2J+L)}{(G-L)(2G-2E+J-L)} \frac{ \partial_i \partial_j \ddot{q}}{\square \nabla^6}\\
\frac{h}{\ks} &=& \frac{q}{(2G-2E+J-L) \nabla^4} - \frac{\rho}{D \nabla^2}
\eea
The coefficients in the exchange action eq.~(\ref{LexB}) are found to be
\bea
&&b_1=-\frac{L \left[-2 E^2+2 E L+L (-2 G-3 J+L)+4 G J\right]}{D^2 (2 E-2 G-J+L)},\,\,\,\,\,\,
b_2=\frac{2 A^2 \left[E^2-2 E L+2 L (G+J)-2 G J-L^2\right]}{D^2 H (2 E-2 G-J+L)}, \,\,\,\,\,\, \nonumber \\
&&b_3 = \frac{2 F L}{F H-H^2},\,\,\,\,\,\, b_4 = \frac{A^2 [(3 F+H) (2 E-2 G-J)+L (F+3 H)]}{2 H^2 (F-H) (2 E-2 G-J+L)} ,\,\,\,\,\,\, b_5 = \frac{-A^4}{H^3 (2 E-2 G-J+L)} \nonumber \\
&& b_6 = \frac{A L (-4 E+6 G+J-3 L)}{D H (2 E-2 G-J+L)} ,\,\,\,\,\,\, b_7 = \frac{-2 A^3 (E-2 G+L)}{D H^2 (2 E-2 G-J+L)} ,\,\,\,\,\,\, b_8 = \frac{-2 E+2 G+J}{2 E-2 G-J+L}+\frac{2 G}{G-L}-\frac{1}{2} \nonumber \\
&& b_9 = \frac{A^2}{H} \left(\frac{1}{2 E-2 G-J+L}+\frac{2}{G-L}\right) , \,\,\,\,\,\, b_{10} = -\frac{L (4 E-2 G-3 J+L)}{D (2 E-2 G-J+L)} , \,\,\,\,\,\, b_{11} = -\frac{2 A^2 [3 E-2 (G+J)+L]}{D H (2 E-2 G-J+L)} \nonumber \\
&& b_{12} = \frac{A (-2 E+2 G+J-3 L)}{H (2 E-2 G-J+L)} , \,\,\,\,\,\, b_{13} = \frac{-2 A^3}{H^2 (2 E-2 G-J+L)}
\eea

\section{Supplementary Details for Theory C} \label{B}

The form of the exchange action that includes non-local parts in Theory (C) is found to take the following form
\bea
{\Lnonlocal\over\kappa^2} &=&  c_{pp1} \,p_i {\ddot{p}_i \over \square \nabla^2} +  c_{\rho \rho 1} \,\rho {\ddot{\rho} \over \square \nabla^2}  +  c_{\rho \rho 2} \,\rho {\ddddot{\rho} \over \square \nabla^4}   + c_{\rho \tau} \,\rho {\ddot{\tau} \over \square \nabla^2} + c_{\rho f 1}\, \rho {\dddot{f} \over \square  \nabla^2} + c_{\rho f 2}\,\rho {\partial_t^5 f \over \square \nabla^4} 
 + c_{\tau f 1}\, \tau {\dddot{f} \over \square  \nabla^2} \nonumber\\
 &+& c_{\tau f 2}\, \tau {\partial_t^5 f \over \square \nabla^4} +  c_{\rho g 1}\, \rho {\ddot{g} \over \square \nabla^2}  +  c_{\rho g 2}\, \rho {\ddddot{g} \over \square \nabla^4} +  c_{\tau g}\, \tau { \ddot{g} \over \square \nabla^2 } 
+ c_{ff1}\, f {\ddddot{f} \over \square \nabla^2} + c_{ff2}\, f { \partial_t^6 f \over \square \nabla^2} + c_{f g 1}\, f {\dddot{g} \over \square  \nabla^2}\nonumber\\
 &+& c_{ f g 2}\,f {\partial_t^5 g \over \square \nabla^4} + c_{gg1}\, g { \ddot{g}\over \square \nabla^2} + c_{gg2}\,  g { \ddddot{g}  \over \square \nabla^4} 
\eea
The 17 coefficients $c_{pp1},c_{\rho\rho1},\ldots,c_{gg2}$ are relatively complicated functions of the coefficients $A,B,\ldots L$. We do not present the full details of the coefficients here for the sake of brevity.
		
\end{widetext}

\end{document}